\documentclass[letterpaper,12pt,notoc]{JHEP3}
\def\pd{\partial}
\def\mc{\mathcal}

\usepackage{graphicx}
\usepackage{amsmath}
\usepackage{amssymb}

\preprint{ \hbox{}\hfill arXiv:1307.6641}

\title{$\frac{1}{2}$-BPS Domain wall from $N=10$ three
dimensional gauged supergravity}
\author{Parinya Karndumri\\
String Theory and Supergravity Group, Department
of Physics, Faculty of Science, Chulalongkorn University, 254 Phayathai Road, Pathumwan, Bangkok 10330, Thailand\\
Thailand Center of Excellence in Physics, CHE, Ministry of
Education, Bangkok 10400, Thailand
\\
E-mail: \email{parinya.ka@hotmail.com}}

\abstract{We explicitly construct $N=10$ Chern-Simons gaged
supergravity in three dimensions with non-semisimple gauge group
$SO(5)\ltimes \mathbf{T}^{10}$. The gauge group is embedded in
$E_{6(-14)}$ which is the isometry group of the 32-dimensional
scalar manifold $E_{6(-14)}/SO(10)\times U(1)$. The resulting theory
is on-shell equivalent to $SO(5)$ Yang-Mills gauged supergravity
coming from dimensional reduction on $S^1$ of $SO(5)$ $N=5$ gauged
supergravity in four dimensions. We discuss the spectrum of the
corresponding reduction. The $SO(5)\ltimes \mathbf{T}^{10}$ gauged
supergravity, describing the reduced theory, admits a
$\frac{1}{2}$-BPS domain wall vacuum solution whose explicit form is
also given. This provides an example of a domain wall in non-maximal
gauged supergravity.} 
\begin{document}
\section{Introduction}
Chern-Simons gauged supergravity in three dimensions has a very rich
structure due to the duality between scalars and vectors in three
dimensions. There are many possible gauge groups since there is no
restriction on the number of vector fields that act as gauge fields
\cite{nicolai1, N8}, or equivalently, no restriction on the
dimension of the gauge group provided that it can be embedded in the
global symmetry group and consistent with supersymmetry. Any number
of vector fields can be introduced via Chern-Simons terms which do
not give rise to extra degrees of freedom. The theory is also useful
in the study of AdS$_3$/CFT$_2$ correspondence, see for example
\cite{krause lecture} for a nice review.
\\
\indent To understand AdS$_3$/CFT$_2$ correspondence in the context
of string/M theory, the embedding of three dimensional gauged
supergravity in ten or eleven dimensions is required. The usual
procedure to obtain lower dimensional supergravities from higher
dimensional theories is the Kaluza-Klein (KK) dimensional reduction.
The general U-duality covariant formulation of three dimensional
gauged supergravities is in the form of Chern-Simons theory in which
the gauge fields enter the Lagrangian through the Chern-Simons terms
\cite{dewit}. On the other hand, dimensional reductions result in
Yang-Mills type gauged supergravity in which gauge kinetic terms are
in the form of conventional Yang-Mills terms. The known class of
Chern-Simons gauge groups that gives equivalent Yang-Mills type
theory is of non-semisimple type \cite{csym}. Any Yang-Mills type
Lagrangian can be rewritten in the Chern-Simons form by introducing
two gauge fields and a compensating scalar for each Yang-Mills gauge
field. This makes non-semisimple gauge groups more interesting in
finding effective theories of string/M theory in three dimensions.
\\
\indent Some embeddings of three dimensional gauged supergravities
into higher dimensions have appeared so far. These examples include
$N=2,4,8,16$ gauged supergravities from reductions on spheres and
Calabi-Yau manifold in \cite{PopeSU2, PopeSU22, KK, henning_ADSCMT,
nicolai3, Hohm_henning, henning_N8AdS3_S3} and recently various
$N=2$ theories from wrapped D3-branes of \cite{EOC_PKD3}. In this
paper, we will give another example of this embedding namely $N=10$
gauged supergravity with $SO(5)\ltimes \mathbf{T}^{10}$ gauge group.
Due to the above mentioned equivalent between Chern-Simons and
Yang-Mills type gauged supergravities, this should potentially
describe $N=5$ gauged supergravity in four dimensions with gauged
group $SO(5)$ reduced on $S^1$. The latter has been constructed in
\cite{N5_4D}. It has been shown in \cite{N5_4D_vacua} that the
theory admits two $AdS_4$ critical points, an $N=5$ supersymmetric
point with $SO(5)$ gauge symmetry and a non-supersymmetric point
with $SO(3)$ residual gauge symmetry. The theory has also been
studied in the context of holographic superconductor in
\cite{N5_holo_superconductor}. The non-supersymmetric critical point
is perturbatively stable with all mass-squares above the BF-bound.
\\
\indent Unlike the four dimensional analogue which has maximally
supersymmetric $AdS_4$ ground state, we will find that the reduced
theory in three dimensions admits only a $\frac{1}{2}$-BPS domain
wall as a vacuum solution. This is in contrast to compact and
non-compact gaugings of the same theory studied in \cite{AP2} that
admits maximally supersymmetric $AdS_3$ critical points. The loss of
supersymmetry after $S^1$ reduction has been pointed out in the
context of non-semisimple gaugings in three dimensions in
\cite{nicolai3}. A general result on $S^1$ reduction of AdS spaces
has been given in \cite{DW_from_AdS}. There are many known
$\frac{1}{2}$-BPS domain walls in higher dimensional gauged
supergravities, see for example \cite{DW_Hull1, DW_Hull2,
Pope_dilatonic_brane, Eric_DW1, Eric_DW2, eric_susyDW} as well as in
lower dimensions, see \cite{DW3D} and \cite{Henning_2D} for three-
and two-dimensional solutions. These domain walls are important in
the context of the DW/QFT correspondence \cite{DW/QFT_townsend,
correlator_DW/QFT, Skenderis_DW/QFT} which is a generalization to
non-conformal field theories of the original AdS/CFT correspondence
\cite{maldacena}. They are also useful in the study of domain
wall/cosmology \cite{DW_cosmology, DW_cosmology1, DW_cosmology2}.
\\
\indent The paper is organized as follow. In section
\ref{N10_theory}, we review the general structure of $N$ extended
gauged supergravities in three dimensions including all relevant
formulae and notations. The $SO(5)\ltimes \mathbf{T}^{10}$ gauged
supergravity and the associated domain wall solution are discussed
in section \ref{DW_solution}. We then discuss possible higher
dimensional origin of the resulting theory from $S^1$ dimensional
reduction of $N=5$ $SO(5)$ gauged supergravity in four dimensions.
We finally give some conclusions and comments in section
\ref{conclusion}. All details and explicit calculations are given in
appendix \ref{detail}. In appendix \ref{N9_theory}, we will explore
possible non-semisimple gauge groups of $N=9$ gauged supergravity in
three dimensions.
\section{$N=10$ gauged supergravity in three dimensions with non-semisimple gauge
groups}\label{N10_theory} Before going to the detail of the
construction, we briefly review the general structure of three
dimensional gauged supergravities and apply it to the construction
of $N=10$ gauged supergravity with non-semisimple gauge group
$SO(10)\ltimes \mathbf{T}^{10}$. We will keep the number of
supersymmetry to be $N$ for conveniences and later set $N=10$. In
general, the matter coupled supergravity in three dimensions is in
the form of a non-linear sigma model coupled to supergravity. For
$N>4$, supersymmetry demands that the scalar target manifold must be
a symmetric space of the form $G/H$ in which $G$ and $H$ are the
global symmetry group and its maximal compact subgroup, respectively
\cite{dewit1}. In particular, for $N>8$, supersymmetry determines
the scalar manifold uniquely. In the present case of $N=10$, the
scalar manifold is given by the coset space $E_{6(-14)}/SO(10)\times
U(1)$ which is a 32-dimensional Kahler manifold.
\\
\indent Coupling of the sigma model to N-extended supergravity
requires the presence of $N-1$ almost complex structures $f^P$,
$P=2,\ldots, N$ on the scalar manifold. The tensors
$f^{IJ}=f^{[IJ]}$, $I,J=1,\ldots, N$, constructed by the relation
\begin{equation}
f^{1P}=-f^{P1}=f^P,\qquad f^{PQ}=f^{[P}f^{Q]}\, .\label{fIJ}
\end{equation}
generate the $SO(N)$ R-symmetry in a spinor representation under
which scalar fields transform. On symmetric scalar manifolds of the
form $G/H$, the maximal compact subgroup $H=SO(N)\times H'$ contains
the R-symmetry $SO(N)$ and another compact subgroup $H'$ commuting
with $SO(N)$. In $N=10$ theory, the group $H'$ is simply $U(1)$. The
$G$-generators $t^{\mc{M}}$, $\mc{M}=1,\ldots, \textrm{dim}G$, can
be split into $(T^{IJ}, X^\alpha)$ generating, respectively,
$SO(N)\times H'$ and non-compact generators $Y^A$ corresponding to
$\textrm{dim}\, G-\textrm{dim}\, H$ scalars. The global symmetry
group $G$ is characterized by the following algebra
\begin{eqnarray}
\left[T^{IJ},T^{KL}\right]&=&-4\delta^{[I[K}T^{L]J]}, \qquad
\left[T^{IJ},Y^A\right]=-\frac{1}{2}f^{IJ,AB}Y_B, \nonumber \\
\left[X^\alpha,X^\beta\right]&=&f^{\alpha
\beta}_{\phantom{as}\gamma}X^\gamma,\qquad
\left[X^\alpha,Y^A\right]=h^{\alpha
\phantom{a}A}_{\phantom{a}B}Y^B, \nonumber \\
\left[Y^{A},Y^{B}\right]&=&\frac{1}{4}f^{AB}_{IJ}T^{IJ}+\frac{1}{8}C_{\alpha\beta}h^{\beta
AB}X^\alpha\, . \label{Galgebra}
\end{eqnarray}
The tensors $f^{IJ}$ are related to $SO(N)$ gamma matrices,
$\Gamma^I_{A\dot{A}}$ in which $A$ and $\dot{A}$ label spinor and
conjugate spinor representations, respectively, by
\begin{equation}
f^{IJ}=-\frac{1}{2}\Gamma^{IJ}=-\frac{1}{4}\left(\Gamma^I\Gamma^J-\Gamma^J\Gamma^I\right).
\end{equation}
$C_{\alpha\beta}$ and $f^{\alpha \beta}_{\phantom{as}\gamma}$ are
$H'$ invariant tensor and $H'$ structure constants, respectively.
The $H'$ group is generated in the $SO(N)$ spinor representation by
matrices $h^{\alpha \phantom{a}A}_{\phantom{a}B}$. The coset
manifold whose coordinates are given by $d=\textrm{dim}(G/H)$ scalar
fields $\phi^i$, $i=1,\ldots , d$ can be described by a coset
representative $L$. The usual formulae for a coset space are
\begin{eqnarray}
L^{-1}t^\mathcal{M}L&=&\frac{1}{2}\mathcal{V}^{\mathcal{M}}_{\phantom{as}IJ}T^{IJ}+\mathcal{V}^\mathcal{M}_{\phantom{as}\alpha}X^\alpha+
\mathcal{V}^\mathcal{M}_{\phantom{as}A}Y^A,\label{cosetFormula}\\
L^{-1} \partial_i L&=& \frac{1}{2}Q^{IJ}_i T^{IJ}+Q^\alpha_i
X^{\alpha}+e^A_i Y^A \label{cosetFormula1}
\end{eqnarray}
which will be useful later on. $e^A_i$ is the vielbein on the scalar
manifold while $Q^{IJ}_i$ and $Q^\alpha_i$ are $SO(N)\times H'$
composite connections. Scalar matrices $\mc{V}$ will be used to
define the moment maps below.
\\
\indent Gaugings of supergravities in various space-time dimensions
are efficiently described in a $G$-covariant way by the so-called
embedding tensor formalism \cite{nicolai1}. In essence, the
embedding tensor $\Theta_{\mc{MN}}$ is a symmetric gauge invariant
tensor that acts as a projector from the global symmetry group $G$
to a particular gauge group. Gauge covariant derivatives describing
the minimal coupling of the gauge fields $A^\mc{M}_\mu$ to other
fields also involve the embedding tensor. For example, the covariant
derivative on scalar fields is given by
\begin{equation}
\mc{D}_\mu\phi^i=\pd_\mu \phi^i+g \Theta_{\mc{MN}}A^\mc{M}_\mu
X^{\mc{N}i}
\end{equation}
where $X^{\mc{N}i}$ are Killing vectors generating isometries on the
scalar manifold and $g$ is the gauge coupling constant.
\\
\indent In order to define a viable gauging, the embedding tensor
has to satisfy the so-called quadratic constraint
\begin{equation}
\Theta_{\mathcal{PL}}f^{\mathcal{KL}}_{\phantom{asds}\mathcal{(M}}\Theta_{\mathcal{N)K}}=0,\label{theta_quadratic}
\end{equation}
which is the requirement that the gauge generators
$\Theta_{\mc{MN}}t^\mc{N}$ form a closed algebra, or equivalently
the gauge group is a proper subgroup of $G$. Furthermore, for
supersymmetry to be preserved in the gauging process, the embedding
tensor needs to satisfy the projection constraint
\begin{equation}
\mathbb{P}_{R_0}\Theta_{\mc{MN}}=0\, .\label{theta_projection}
\end{equation}
This condition comes from supersymmetry, but it should be noted that
the constraint in this form is obtained by regarding the scalar
manifold to be a symmetric space.
\\
\indent It is useful to introduce the T-tensor given by the moment
map of the embedding tensor by scalar matrices
$\mc{V}^\mc{M}_{\phantom{as}\mc{A}}$, obtained from
\eqref{cosetFormula},
\begin{equation}
T_{\mathcal{AB}}=\mathcal{V}^{\mc{M}}_{\phantom{as}\mc{A}}\Theta_{\mc{MN}}\mathcal{V}^{\mc{N}}_{\phantom{as}\mc{B}}\,
.\label{T_tensor_def}
\end{equation}
The T-tensor transforms under the maximal compact subgroup $H$ and
consists of various components such as $T^{IJ,KL}$, $T^{IJ,A}$ and
$T^{A,B}$. Since fermions transform under $H$, the fermion couplings
will be written in term of the T-tensor or linear combinations of
its components as we will see below. For any supersymmetric gauging,
supersymmetry requires only that the T-tensor satisfies the
projection
\begin{equation}
\mathbb{P}_\boxplus T^{IJ,KL}=0\label{Tconstraint1}
\end{equation}
where $\boxplus$ is the Riemann tensor-like representation of
$SO(N)$. In the case of symmetric scalar manifolds which are of
interest in this paper, this constraint can be lifted to the
constraint on the embedding tensor given in \eqref{theta_projection}
in which the $G$-representation $R_0$, branched under $SO(N)$,
contains $\boxplus$ representation of $SO(N)$. Any subgroup of $G$
whose embedding tensor satisfies the above constraints is called
admissible gauge group.
\\
\indent In general, gaugings need some modifications to the original
ungauged Lagrangian by fermionic mass-like terms and a scalar
potential, at order $g$ and $g^2$, respectively. Also, the
supersymmetry transformation rules need to be modified at order $g$.
In what follow, we will need the scalar potential and fermionic
supersymmetry transformations. They are written in terms of the
$A_1^{IJ}$ and $A_{2i}^{IJ}$ tensors which are in turn constructed
from various components of the T-tensor
\begin{eqnarray}
A_1^{IJ}&=&-\frac{4}{N-2}T^{IM,JM}+\frac{2}{(N-1)(N-2)}\delta^{IJ}T^{MN,MN},\label{A1}\\
A_{2j}^{IJ}&=&\frac{2}{N}T^{IJ}_{\phantom{as}j}+\frac{4}{N(N-2)}f^{M(I
m}_{\phantom{as}j}T^{J)M}_{\phantom{as}m}+\frac{2}{N(N-1)(N-2)}\delta^{IJ}f^{KL\phantom{a}m}_{\phantom{as}j}T^{KL}_{\phantom{as}m}\,
.\label{A2}\nonumber \\
& &
\end{eqnarray}
The scalar potential is simply given by
\begin{equation}
V=-\frac{4}{N}g^2\left(A_1^{IJ}A_1^{IJ}-\frac{1}{2}Ng^{ij}A_{2i}^{IJ}A_{2j}^{IJ}\right).\label{potential}
\end{equation}
The metric $g_{ij}$ on the target manifold is related to the
vielbein by $g_{ij}=e^A_{i}e^A_j$. We also note here that the
quadratic constraint \eqref{theta_quadratic} can be written in terms
of $A_1^{IJ}$ and $A_{2i}^{IJ}$ as
\begin{equation}
2A_1^{IK}A_1^{KJ}-NA_2^{IKi}A_{2i}^{JK}=\frac{1}{N}\delta^{IJ}\left(2A_1^{KL}A_1^{KL}-NA_2^{KLi}A^{KL}_{2i}\right).\label{quadratic2}
\end{equation}
\indent The fermionic field content of the $N$ extended supergravity
in three dimensions consists of $N$ gravitini $\psi^I_\mu$ and $d$
spin-$\frac{1}{2}$ fields $\chi^{iI}$. The latter is written in an
overcomplete basis and subject to the projection constraint
\begin{equation}
\chi^{iI}=\frac{1}{N}\left(\delta^{IJ}\delta^i_j-f^{IJi}_{\phantom{asc}j}\right)\chi^{jJ}
\end{equation}
giving rise to $d$ independent $\chi^{iI}$ fields. The fermions
$\chi^{iI}$ can be redefined such that they transform in a conjugate
spinor representation of $SO(N)$ via \cite{dewit}
\begin{equation}
\chi^{\dot{A}}=\frac{1}{N}e^A_i\Gamma^I_{A\dot{A}}\chi^{iI}\, .
\end{equation}
The corresponding supersymmetry transformations are as follow:
\begin{eqnarray}
\delta\psi^I_\mu
&=&\mathcal{D}_\mu\epsilon^I+gA_1^{IJ}\gamma_\mu\epsilon^J,\label{d_psi}\\
\delta\chi^{iI}&=&
\frac{1}{2}(\delta^{IJ}\mathbf{1}-f^{IJ})^i_{\phantom{a}j}{\mathcal{D}{\!\!\!\!/}}\phi^j\epsilon^J
-gNA_2^{JIi}\epsilon^J\label{d_chi}
\end{eqnarray}
where only relevant terms are given and
\begin{equation}
\mc{D}_\mu
\epsilon^I=\pd_\mu\epsilon^I+\frac{1}{4}\omega_\mu^{ab}\gamma_{ab}+\pd_\mu\phi
Q_i^{IJ}\epsilon^I+g\Theta_{\mc{MN}}A^\mc{M}_\mu
\mc{V}^{\mc{N}IJ}\epsilon^J\, .
\end{equation}
\indent Gauge groups of interest to us are non-semisimple groups of
the form $G_0\ltimes \mathbf{T}^{\textrm{dim}\, G}$. The
translational symmetry $\mathbf{T}^{\textrm{dim}\, G}$ consists of
$\textrm{dim}\, G$ commuting generators which transform as an
adjoint representation under $G_0$. This type of gauge groups gives
rise to the on-shell equivalent Yang-Mills gauged supergravity
coming from dimensional reductions of some higher dimensional
theory. The $G_0\ltimes \mathbf{T}^{\textrm{dim}\, G}$ gauge group
whose generators are respectively $J^m$ and $T^m$, $m=1,\ldots,
\textrm{dim}\, G$ is characterized by the following algebra
\begin{eqnarray}
\left[J^m,J^n\right]&=&f^{mn}_{\phantom{ass}k}J^k,\qquad
\left[J^m,T^n\right]=f^{mn}_{\phantom{ass}k}T^k,\qquad
\left[T^m,T^n\right]=0\label{Gauge_algebra}
\end{eqnarray}
where $f^{mn}_{\phantom{ass}k}$ are $G_0$ structure constants. We
will denote the $G_0$ and $\mathbf{T}^{\textrm{dim}\, G}$ parts of
the gauge group by ${\rm{a}}$ and ${\rm{b}}$, respectively. As shown
in \cite{csym}, the corresponding embedding tensor consists of two
parts, one with the coupling between ${\rm{a}}$ and ${\rm{b}}$ types
$\Theta_{{\rm{a}}{\rm{b}}}$ and the other with the coupling between
${\rm{b}}$ and ${\rm{b}}$ types $\Theta_{{\rm{b}}{\rm{b}}}$. The
full embedding tensor can be written as
\begin{equation}
\Theta =g_1\Theta_{{\rm{a}}{\rm{b}}}+g_2\Theta_{{\rm{b}}{\rm{b}}}
\end{equation}
with $g_1$ and $g_2$ being the coupling constants. Supersymmetry
constraint \eqref{theta_projection} may impose some relation on
$g_1$ and $g_2$ such that eventually there is only one coupling.
Both $\Theta_{{\rm{a}}{\rm{b}}}$ and $\Theta_{{\rm{b}}{\rm{b}}}$ are
given by the Cartan-Killing form of $G_0$, $\eta^{G_0}_{mn}$, which
is non-degenerate since $G_0$ is semisimple. The above information
is sufficient for our discussion in this paper. The interested
readers are invited to consult \cite{dewit} and \cite{csym} for more
a detailed discussion about three dimensional gauged supergravity
with non-semisimple gauge groups.
\section{$SO(5)\ltimes \mathbf{T}^{10}$ gauged supergravity and $\frac{1}{2}$-BPS domain wall
solution}\label{DW_solution} In this section, we explicitly
construct $N=10$ gauged supergravity with $SO(5)\ltimes
\mathbf{T}^{10}$ gauge group. We begin with the scalar manifold
$E_{6(-14)}/SO(10)\times U(1)$ and use $E_6$ generators given in
\cite{F4} and \cite{E6}. The non-compact form $E_{6(-14)}$ is
constructed by using the ``Weyl unitarity trick''. We follow the
same construction and notation as in \cite{AP2} to which we refer
the readers for more details.
\\
\indent The 78 generators of $E_6$ constructed in \cite{E6} are
labeled by $c_i$, $i=1,\ldots, 78$. The $SO(10)$ R-symmetry is
generated by $c_i$, $i=1,\ldots, 21, 30,\ldots 36, 45,\ldots, 52,
71,\ldots , 78$ and $\tilde{c}_{53}$. We need to relabel these
generators to the form of $T^{IJ}$ in our $SO(N)$ covariant
formalism. This has already been done in \cite{AP2}, but we will
repeat it in appendix \ref{detail} for convenience. The group
$H'=U(1)$ is generated by $\tilde{c}_{70}$ whose definition and that
of $\tilde{c}_{53}$ can be found in appendix \ref{detail}.
\\
\indent The non-compact generators can be identified as
\begin{equation}
Y^A=\left\{
\begin{array}{rl}
ic_{A+21} &\textrm{for} \; A=1, \ldots, 8\\
ic_{A+28} &\textrm{for} \; A=9, \ldots, 16\\
ic_{A+37}&\textrm{for} \; A=17, \ldots, 32
\end{array}\right. \, .\label{non_compact_gen}
\end{equation}
We can then use \eqref{Galgebra} to extract the tensors $f^{IJ}$
whose components are computed by
\begin{equation}
f^{IJ}_{AB}=-\frac{1}{3}\textrm{Tr}\left(\left[T^{IJ},Y^A\right]Y^B\right).
\end{equation}
Notice that the generators have normalizations
$\textrm{Tr}(T^{IJ}T^{IJ})=-6$ and $\textrm{Tr}(Y^{A}Y^{A})=6$, no
sum on $IJ$ and $A$.
\\
\indent We now construct generators of the gauge group $SO(5)\ltimes
\mathbf{T}^{10}$. This group is embedded in $USp(4,4)\subset
E_{6(-14)}$. The maximal compact subgroup $USp(4)\times
USp(4)\subset USp(4,4)$ is identified as the $SO(5)\times SO(5)$
subgroup of the R-symmetry $SO(10)$. Recall that the 32 scalars
transform as $\mathbf{16}^++\mathbf{16}^-$ under $SO(10)\times
U(1)$. Under $SO(5)\times SO(5)$, the scalars transform as
\begin{equation}
\mathbf{16}^++\mathbf{16}^-=(\mathbf{4},\mathbf{4})^++(\mathbf{4},\mathbf{4})^-\,
.
\end{equation}
We then identify $SO(5)$ part of the gauge group as the diagonal
subgroup $SO(5)_{\textrm{diag}}\subset SO(5)\times SO(5)$ under
which scalars transform as
\begin{eqnarray}
\mathbf{16}^++\mathbf{16}^-&=&(\mathbf{4}\times\mathbf{4})^++(\mathbf{4}\times\mathbf{4})^-\nonumber \\
&=&(\mathbf{1}+\mathbf{10}+\mathbf{5})^++(\mathbf{1}+\mathbf{10}+\mathbf{5})^-\,
.\label{N10scalar_decom}
\end{eqnarray}
In this decomposition, we see that there are two singlets under
$SO(5)_{\textrm{diag}}$. The adjoint representation $\mathbf{10}^+$
and $\mathbf{10}^-$ will be used to construct the translational
generators of $\mathbf{T}^{10}$.
\\
\indent The explicit form of the corresponding gauge generators are
as follow. The $SO(5)_{\textrm{diag}}$ generators are given by
\begin{equation}
J^{ij}=T^{ij}+T^{i+5,j+9},\qquad i,j=1,\ldots, 5
\end{equation}
while the $\mathbf{T}^{10}$ generators are found to be
\begin{equation}
t^{ij}=T^{ij}-T^{i+5,j+5}+\tilde{Y}^{ij},\qquad ,i,j=1,\ldots, 5
\end{equation}
where $\tilde{Y}^{ij}$ are given in appendix \ref{detail}.
\\
\indent The embedding tensor is of the form
\begin{equation}
\Theta=g_1 \Theta_{{\rm{a}}{\rm{b}}}+g_2 \Theta_{{\rm{b}}{\rm{b}}}
\end{equation}
where $\Theta_{{\rm{a}}{\rm{b}}}$ and $\Theta_{{\rm{b}}{\rm{b}}}$
are given by the Cartan-Killing form of $SO(5)$. The supersymmetry
constraint requires $g_2=0$ meaning that there is no coupling among
$\mathbf{T}^{10}$ generators. This is similar to $N=16$ and $N=8$
theories with $SO(8)\ltimes \mathbf{T}^{28}$ gauge group studied in
\cite{nicolai3, DW3D}.
\\
\indent We are now in a position to study the scalar potential of
the resulting gauged supergravity. Following the technique of
\cite{warner}, we begin with scalar fields which are singlets under
the semisimple part of the gauge group, $SO(5)$. They are given by
$\mathbf{1}^\pm$ in \eqref{N10scalar_decom} and correspond to the
non-compact generators
\begin{eqnarray}
Y_{s1}&=&Y_3-Y_5-Y_{12}+Y_{16}+Y_{17}-Y_{18}+Y_{27}+Y_{29},\nonumber
\\
Y_{s2}&=&Y_4+Y_8+Y_{11}+Y_{13}+Y_{22}-Y_{23}+Y_{28}-Y_{32}\, .
\end{eqnarray}
Accordingly, the coset representative is parametrized by
\begin{equation}
L=e^{a Y_{s1}}e^{b Y_{s2}}\, .
\end{equation}
Using the formulae \eqref{V_map} and \eqref{T_ten}, we can compute
$A_1^{IJ}$ and $A_{2i}^{IJ}$ by using a computer program
\textsl{Mathematica}. The scalar potential is computed to be
\begin{equation}
V=-6e^{4(a-b)}\left(1+e^{8b}\right)g^2
\end{equation}
where we have denoted $g_1$ simply by $g$. The presence of the $e^a$
factor implies that the potential has no critical point. We then
expect the vacuum solution to be a domain wall.
\\
\indent To find a domain wall solution, we adopt the usual domain
wall ansatz for the metric
\begin{equation}
ds^2=e^{2A}dx^2_{1,1}+dr^2\, .
\end{equation}
The supersymmetry transformation of $\chi^{iI}$, $\delta
\chi^{iI}=0$ from equation \eqref{d_chi}, gives the following
equations
\begin{eqnarray}
b'\gamma_r\epsilon^I+\frac{1}{2}g(1-e^{4b})e^{2(a-b)}\epsilon^I&=&0,\qquad I=1,\ldots ,5,\label{eq11}\\
b'\gamma_r\epsilon^I-\frac{1}{2}g(1-e^{4b})e^{2(a-b)}\epsilon^I&=&0,\qquad
I=6,\ldots ,10, \label{eq12}\\
a'\gamma_r\epsilon^I-g\frac{e^{2(a+b)\left(1+e^{4b}\right)}}{1+e^{8b}}\epsilon^I&=&0,\qquad I=1,\ldots ,5,\label{eq21}\\
a'\gamma_r\epsilon^I+g\frac{e^{2(a+b)\left(1+e^{4b}\right)}}{1+e^{8b}}\epsilon^I&=&0,\qquad
I=6,\ldots ,10\label{eq22}
\end{eqnarray}
where we have used $'$ to denote the derivative $\frac{d}{dr}$ and
${\phi^A}'=\frac{1}{6}\textrm{Tr}\left(L^{-1}L'Y^A\right)$. We will
now impose the projection conditions
$\gamma_r\epsilon^I=-\epsilon^I$ for $I=1,\ldots, 5$ and
$\gamma_r\epsilon^I=\epsilon^I$ for $I=6,\ldots, 10$. $\epsilon^I$
has two real components. The projectors then reduce the
supersymmetry by a fraction of $\frac{1}{2}$. With these two
projectors, we end up with two independent equations
\begin{eqnarray}
b'&=&\frac{1}{2}g(1-e^{4b})e^{2(a-b)},\label{eq1}\\
a'&=&-g\frac{e^{2(a+b)\left(1+e^{4b}\right)}}{1+e^{8b}}\,
.\label{eq2}
\end{eqnarray}
The supersymmetry variation of the gravitini $\psi^I_\mu$, $\delta
\psi^I_\mu=0$ from equation \eqref{d_psi} after using the above
projectors, gives rise to
\begin{eqnarray}
e^{4b}&=&1,\label{eq3}\\
A'&=&2g\left(1+e^{4b}\right)e^{2(a-b)}\label{eq4}
\end{eqnarray}
where we have used the spin connection
$\omega_{\hat{\mu}}^{\hat{\nu}\hat{r}}=A'\delta^{\hat{\nu}}_{\hat{\mu}}$
with $\hat{\mu}, \hat{\nu}=0,1$.
\\
\indent We see from \eqref{eq3} that supersymmetry demands $b=0$.
Equation \eqref{eq1} is now trivially satisfied, and equation
\eqref{eq2} becomes
\begin{equation}
a'+e^{2a}g=0\, .
\end{equation}
The solution is easily obtained to be
\begin{equation}
a=-\frac{1}{2}\ln \left(2gr+C_1\right)
\end{equation}
where $C_1$ is an integration constant. Substituting into equation
\eqref{eq4} gives
\begin{equation}
A'=4ge^{2a}=\frac{4g}{C_1+2gr}
\end{equation}
whose solution is, with another integration constant $C_2$,
\begin{equation}
A=C_2+2\ln\left(2gr+C_1\right).
\end{equation}
\indent As in other solutions of this type, the residual
supersymmetry is generated by the Killing spinors given by
$\epsilon^i=e^{\frac{A}{2}}\epsilon^i_{0\pm}$, $i=1,\ldots ,5$ with
the constant spinors $\epsilon^i_{0\pm}$ satisfying
$\gamma_r\epsilon^i_{0\pm}=\pm\epsilon^i_{0\pm}$. The full symmetry
of this solution is $ISO(1,1)\times SO(5)$ with the unbroken
$N=(5,5)$ Poincare supersymmetry in notation of the dual
two-dimensional field theory.
\\
\indent The two integration constants $C_1$ and $C_2$ can be set to
zero by shifting the coordinate $r$ and rescaling the coordinates
$x^\mu$. We can also write down the solution in the form of warped
$AdS_3$ by introducing the new coordinate $\rho$ via
$\rho=-\frac{1}{4g^2r}$ in term of which the metric becomes
\begin{equation}
ds^2=\frac{1}{\left(4g^2\rho\right)^2}\left(\frac{dx^2_{1,1}+d\rho^2}{\rho^2}\right).
\end{equation}
\indent We end this section by considering subgroups of
$SO(5)\ltimes \mathbf{T}^{10}$ namely $SO(4)\ltimes \mathbf{T}^6$
and $(SO(3)\ltimes \mathbf{T}^3)\times (SO(2)\ltimes
\mathbf{T}^1)\sim U(2)\ltimes \mathbf{T}^4$. It can be checked that
both of them are not admissible.
\section{Higher dimensional origin}\label{Reduction}
In this section, we discuss higher dimensional origin of the
$SO(5)\ltimes \mathbf{T}^{10}$ $N=10$ gauged supergravity
constructed in the previous section. By the general result of
\cite{csym}, this theory is on-shell equivalent to the $SO(5)$
Yang-Mills gauged supergravity which can be obtained from $S^1$
reduction of $N=5$ gauged supergravity in four dimensions with
$SO(5)$ gauge group. The four dimensional theory has been
constructed in \cite{N5_4D} and can be obtained as a truncation of
the maximal $N=8$ gauged supergravity. In the notation of
\cite{N5_4D}, the field content of this theory contains one graviton
$e^a_M$ or $g_{MN}$, five gravitini $\psi^i_M$, eleven
spin-$\frac{1}{2}$ fields $\chi^{ijk}$ and $\chi^{678}$, ten scalars
$\phi^i$ and $\phi_i$ living in the coset space $SU(5,1)/U(5)$ and
ten vector fields $A^{ij}_M$ being $SO(5)$ gauge fields. Here,
$M,N=0,1,2,3$ and $a,b=0,1,2,3$ are four dimensional space-time and
tangent space indices respectively while $i,j=1,\ldots, 5$ are
$SU(5)$ indices except for $A^{ij}_M$ which transform in the adjoint
representation of $SO(5)$.
\\
\indent If we reduce this theory on $S^1$ along the $x^3$ direction,
we find the following fields in three dimensions. The metric
$g_{MN}$ gives the non-dynamical three dimensional metric
$g_{\mu\nu}$, the graviphoton $g_{\mu 3}$ and a scalar $g_{33}$. The
$SO(5)$ gauge fields result in the three dimensional gauge fields of
the same gauge group $A^{ij}_\mu$ and ten scalars $A^{ij}_3$
transforming in the adjoint representation of $SO(5)$. Finally, the
ten scalars $(\phi^i,\phi_i)$ obviously become the three dimensional
scalars.
\\
\indent A spinor in four dimensions give rise to two spinors in
three dimensions. We then obtain ten gravitini $\psi^i_\mu$ from
$\psi^i_M$ and ten spin-$\frac{1}{2}$ fields $\psi^i_3$. There are
additional $20+2$ spin-$\frac{1}{2}$ fields from the reduction of
$\chi^{ijk}$ and $\chi^{678}$, respectively. In three dimensions,
the metric and gravitini do not have any dynamics. We then find 32
fernionic on-shell degrees of freedom from
$(\psi^i_3,\chi^{678},\chi^{ijk})$. We can also dualize $A^{ij}_\mu$
and $g_{\mu 3}$ to $10+1$ scalars. All together, we end up with 32
scalars from $(\phi^i,\phi_i,g_{33},g_{\mu 3},A^{ij}_\mu,A^{ij}_3)$.
This is the same as in $N=10$ gauged supergravity.
\\
\indent We give $SO(5)_{\textrm{gauge}}$ representations of the
reduced fields in table \ref{table1} from which we have omitted the
non-dynamical fields $g_{\mu\nu}$ and $\psi^i_\mu$.
\TABLE{\begin{tabular}{|c|c|c|}
               \hline
               3D fields & $SO(5)$ representation & number of degrees of freedom
               \\ \hline
               $g_{33}$ & $\mathbf{1}$ & 1 \\
               $g_{\mu 3}$ & $\mathbf{1}$ & 1 \\
               $\phi^i$ & $\mathbf{5}$ & 5 \\
               $\phi_i$ & $\mathbf{5}$ & 5 \\
               $A^{ij}_\mu$ & $\mathbf{10}$ & 10 \\
               $A^{ij}_{3}$ & $\mathbf{10}$ & 10 \\
               $\psi^i_3$ & $\mathbf{5}$ & 10 \\
               $\chi^{678}$ & $\mathbf{1}$ & 2 \\
                $\chi^{ijk}$ & $\mathbf{10}$ & 20 \\
               \hline
\end{tabular}\caption{Representations of three dimensional fields resulted from
$S^1$ reduction of $N=5$ gauged supergravity in four dimensions.
}}\label{table1} We have kept $\phi^i$ and $\phi_i$ separately to
emphasize their four dimensional origin. We now consider the
representation of the 32 scalars in $E_{6(-14)}/SO(10)\times U(1)$
coset space under the $SO(5)$ part of the gauge group. Recall that
under $SO(10)\times U(1)$, the scalars transform as
$\mathbf{16}^++\mathbf{16}^-$. Under $SO(10)\times U(1) \supset
SU(5)\times \overline{U(1)}\times U(1) \supset SO(5)$ in which the
$\overline{U(1)}$ is the $U(1)$ subgroup of $U(5)\subset SO(10)$, we
find
\begin{eqnarray}
\mathbf{16}^++\mathbf{16}^-&
&\rightarrow(\mathbf{1}_{-5}+\bar{\mathbf{5}}_3+\mathbf{10}_{-1})^++(\mathbf{1}_{-5}+\mathbf{5}_{-3}
+\overline{\mathbf{10}}_{1})^-\nonumber
\\ & & \rightarrow(\mathbf{1}+\mathbf{5}+\mathbf{10})+(\mathbf{1}+\mathbf{5}+\mathbf{10})
\end{eqnarray}
We find perfect agreement with table \ref{table1}. Reference
\cite{Slansky} is very useful in this decomposition. In the
formalism of $\cite{dewit}$, the fermions $\chi^{\dot{A}}$ transform
as $\overline{\mathbf{10}}^++\mathbf{10}^-$ under $SO(10)\times
U(1)$. Similar decomposition gives $2\times
(\mathbf{1}+\mathbf{5}+\mathbf{10})$ under $SO(5)$ gauge group. This
is again the representations obtained from $S^1$ reduction shown in
table \ref{table1}. The result of \cite{Pope_3D_coset} suggests that
three dimensional supergravity with $E_6$ coset manifold can be
obtained from dimensional reduction on a torus, $S^1$ in the present
case, of a supergravity theory with $A_5$ coset manifold in four
dimensions. Reference \cite{Pope_3D_coset} consider only maximally
non-compact $E_6$ and other types Lie groups. The result here should
provide an example of a non-maximally non-compact $E_6$
$(E_{6(-14)})$ coset obtained from a non-maximally non-compact $A_5$
$SU(5,1)$ coset in four dimensions. Furthermore, the general
formulae for toroidal reductions given in the appendix of
\cite{Pope_3D_coset} should also be applicable in this case.
\section{Conclusions and discussions}\label{conclusion}
In this paper, we have constructed $N=10$ $SO(5)\ltimes
\mathbf{T}^{10}$ gauged supergravity in three dimensions. We have
found that the resulting theory admits a $\frac{1}{2}$-BPS domain
wall as a vacuum solution. The solutions preserves $N=(5,5)$
Poincare supersymmetry in two dimensions with ten supercharges. The
solution is similar to the domain wall from the $S^7$
compactification of type II string theory discussed in
\cite{Henning_Morales}. This solution is the vacuum solution of the
maximal $N=16$ $SO(8)\ltimes \mathbf{T}^{28}$ gauged supergravity.
The solution given here provides an example of a domain wall in
non-maximal gauged supergravity and might be useful in the DW/QFT
correspondence as well as its applications.
\\
\indent We have also discussed possible higher dimensional origin of
this theory. This is given by $S^1$ reduction of $N=5$ $SO(5)$
gauged supergravity in four dimensions. We have found that the
spectrum of the reduction matches with the constructed three
dimensional gauged supergravity. If the $N=5$ four dimensional
theory is reduced on $S^1/\mathbb{Z}_2$, it could give rise to $N=5$
gauged supergravity in three dimensions. Indeed, the latter in
general has scalar manifold $USp(4,k)/USp(4)\times USp(k)$
\cite{dewit1}. We have seen that the $SO(5)\ltimes \mathbf{T}^{10}$
gauge group is embedded in $USp(4,4)\subset E_{6(-14)}$. We then
expect that $N=5$ $SO(5)$ gauged supergravity in four dimensions
reduced on $S^1/\mathbb{Z}_2$ should give $N=5$ $SO(5)\ltimes
\mathbf{T}^{10}$ gauged supergravity in three dimensions with scalar
manifold $USp(4,4)/USp(4)\times USp(4)$ containing 16 scalars. It
turns out that the latter theory admits $SO(5)\ltimes
\mathbf{T}^{10}$ gauge group. The details will be reported in
subsequent work \cite{N5}. Unlike the $N=10$ theory, the $N=5$
truncation admits maximally supersymmetric $AdS_3$ vacuum solution.
This truncation should be similar to the case of $N=8$ $SO(8)\ltimes
\mathbf{T}^{28}$ gauged supergravity with $SO(8,8)/SO(8)\times
SO(8)$ scalar manifold studied in \cite{DW3D}. This theory is a
truncation of $N=16$ $SO(8)\ltimes \mathbf{T}^{28}$ gauged
supergravity with scalar manifold $E_{8(8)}/SO(16)$.
\\
\indent Due to the similar structure as in the above examples, we
would like to briefly discuss the case of $N=12$ gauged
supergravity. The scalar manifold is the 64-dimensional quaternionic
manifold $E_{7(-5)}/SO(12)\times SU(2)$. The gauge group should be
$SO(6)\ltimes \mathbf{T}^{15}$ embedded in $SU(4,4)\subset
E_{7(-5)}$. The $SO(6)$ is again identified as
$SO(6)_{\textrm{diag}}\subset SO(6)\times SO(6)\subset SO(12)$. The
64 scalars transform under $SO(12)\times SU(2)$ as
$(\mathbf{32},\mathbf{2})$ and under $SO(6)\times SO(6)\times SU(2)$
as
$((\mathbf{4},\bar{\mathbf{4}})+(\mathbf{4},\bar{\mathbf{4}}),\mathbf{2})$.
Then, under the $SO(6)$ part of the gauge group, we find the
representation for scalars
$((\mathbf{4}\times\bar{\mathbf{4}}+\mathbf{4}\times\bar{\mathbf{4}}),\mathbf{2})=
(\mathbf{1}+\mathbf{15}+\mathbf{1}+\mathbf{15},\mathbf{2})$. The
non-compact generators in the $\mathbf{15}$ should combine with
$SO(6)\times SO(6)$ generators to form the $\mathbf{T}^{15}$ part of
the gauge group. The fermions transform as
$(\overline{\mathbf{32}},\mathbf{2})$ under $SO(12)\times SU(2)$ and
$((\mathbf{4},\mathbf{4})+(\bar{\mathbf{4}},\bar{\mathbf{4}}),\mathbf{2})$
under $SO(6)\times SO(6)\times SU(2)$. Under $SO(6)$, they transform
as $(\mathbf{10}+\mathbf{6}+\mathbf{10}+\mathbf{6},\mathbf{2})$.
\\
\indent We now consider $S^1$ reduction of $N=6$ $SO(6)$ gauged
supergravity in four dimneions which is also a truncation of $N=8$
$SO(8)$ gauged supergravity \cite{N6_4D}. The bosonic fields are
$(g_{MN}, \phi^{AB},\phi_{AB},A_M^{AB}, A_M)$ where the 30 scalars
$(\phi^{AB},\phi_{AB})$ live in the coset space $SO^*(12)/U(6)$ and
$A,B=1,\ldots, 6$, see \cite{N6_4D} for more detail. The fermionic
fields are given by $(\psi^A_M, \chi^A,\chi^{ABC})$. After $S^1$
reduction, the dynamical bosonic fields are given by $(g_{\mu
3},g_{33},\phi^{AB},\phi_{AB},A_{\mu},A_3,A^{AB}_\mu,A^{AB}_3)$
transforming as
$(\mathbf{1}+\mathbf{1}+\mathbf{15}+\mathbf{15}+\mathbf{1}+\mathbf{1}+\mathbf{15}+\mathbf{15})$
under $SO(6)$ gauge group. After dualizing the vector fields, we end
up with 64 scalars with correct $SO(6)$ representations as in $N=12$
gauged supergravity. The reduced dynamical fermionic fields are
$(\psi^A_3,\chi^{ABC},\chi^A)$ transforming under $SO(6)$ as
$2\times (\mathbf{6}+\mathbf{10}+\mathbf{10}+\mathbf{6})$ which are
indeed the same as those in $N=12$ theory. The factor of 2 comes
from the fact that a four dimensional spinor gives two three
dimensional spinors.
\\
\indent Finally, similar to the discussion in the $N=5$ case, we
expect that the $S^1/\mathbb{Z}_2$ reduction should give $N=6$
$SO(6)\ltimes \mathbf{T}^{15}$ gauged supergravity on three
dimensions with scalar manifold $SU(4,4)/S(U(4)\times U(4))$ whose
compact and non-compact gauge groups have been explored in
\cite{N6}. The possibility of non-semisimple gauge groups is under
investigation \cite{N5}.
\acknowledgments The author would like to thank the Institute for
Fundamental Study (IF), Naresuan University for hospitality during
this work was in progress. He is also grateful to Pitayuth Wongjun
for computing facilities, financial support and many helps during
his visit to IF. This work is partially supported by Thailand Center
of Excellence in Physics through the ThEP/CU/2-RE3/11 project,
Chulalongkorn University through Ratchadapisek Sompote Endowment
Fund and The Thailand Research Fund (TRF) under grant TRG5680010.

\appendix
\section{Useful formulae and details}\label{detail}
In this appendix, we give some details of $N=10$ gauged supergravity
with $SO(5)\ltimes \mathbf{T}^{10}$ gauge group constructed in the
main text. First of all, the $SO(10)$ R-symmetry generators $T^{IJ}$
are explicitly given by
\begin{eqnarray}
T^{1 2}&=& c_1,\,\, \, T^{1 3}= -c_2,\,\, \, T^{2 3}= c_3,\,\, \,
T^{3 4}=
c_6,\,\, \, T^{1 4}= c_4,\,\, \, T^{2 4}= -c_5,\nonumber \\
T^{1 5}&=& c_7,\,\, \, T^{2 5}= -c_8,\,\,\, T^{3 5}= c_9,\,\, \,
T^{4 5}= -c_{10},\,\, \, T^{5 6}= -c_{15},\,\, \, T^{1 6}= c_{11},
\nonumber \\
 T^{2  6}&=&
-c_{12},\,\, \, T^{4 6}= -c_{14},\,\, \, T^{3 6}= c_{13},\,\,\, T^{1
7}= c_{16},\,\, \, T^{27}= -c_{17},\,\, \, T^{4 7}= -c_{19}, \nonumber \\
T^{3 7}&=& c_{18},\,\, \, T^{6 7}= -c_{21},\,\, \, T^{ 5 7}=
-c_{20},\,\, \, T^{7 8}= -c_{36},\,\,\, T^{1 8}= c_{30},\,\, \, T^{2
8}= -c_{31},
\nonumber \\
T^{4 8}&=& -c_{33},\,\, \, T^{3 8}= c_{32},\,\, \, T^{6 8}=
-c_{35},\,\, \, T^{5 8}= -c_{34},\,\, \, T^{2 9}= -c_{46},\,\,\,
T^{1 9}= c_{45},
\nonumber \\
T^{4 9}&=& -c_{48},\,\,\, T^{3 9}= c_{47},\,\,\, T^{6 9}=
-c_{50},\,\, \, T^{5 9}= -c_{49}, T^{8 9}= -c_{52},\,\, \, T^{7 9}=
-c_{51},\nonumber \\
T^{1,10}&=&-c_{71},\,\,\,T^{2,10}=c_{72},\,\,\,T^{3,10}=-c_{73},\,\,\,T^{4,10}=c_{74},\,\,\,
T^{5,10}=c_{75},\nonumber
\\
T^{6,10}&=&c_{76},\,\,\,T^{7,10}=c_{77},\,\,\,T^{8,10}=c_{78},\,\,\,T^{9,10}=-\tilde{c}_{53}\label{TIJ_gen}
\end{eqnarray}
where $\tilde{c}_{53}$ and $\tilde{c}_{70}$ are defined by \cite{E6}
\begin{equation}
\tilde{c}_{53}=\frac{1}{2}c_{53}+\frac{\sqrt{3}}{2}c_{70} \qquad
\textrm{and}\qquad
\tilde{c}_{70}=-\frac{\sqrt{3}}{2}c_{53}+\frac{1}{2}c_{70}\, .
\end{equation}
Also, notice a typo in the sign of $T^{9,10}$ in \cite{AP2}.
\\
\indent The $\tilde{Y}^{ij}$ part of the translational generators
$\mathbf{T}^{10}$ is constructed from the following non-compact
generators
\begin{eqnarray}
\tilde{Y}^{12}&=&\frac{1}{2}\left(Y_3-Y_{12}+Y_{17}+Y_{29}+Y_5-Y_{16}+Y_{18}-Y_{27}\right),\nonumber
\\
\tilde{Y}^{13}&=&\frac{1}{2}\left(Y_2+Y_{14}+Y_{21}-Y_{26}-Y_1+Y_{15}-Y_{19}-Y_{25}\right),\nonumber
\\
\tilde{Y}^{14}&=&\frac{1}{2}\left(Y_{31}-Y_{7}-Y_{6}-Y_{30}-Y_9+Y_{10}+Y_{20}-Y_{24}\right),\nonumber
\\
\tilde{Y}^{15}&=&\frac{1}{2}\left(Y_{15}-Y_{14}+Y_{25}-Y_{26}-Y_1-Y_{2}+Y_{19}+Y_{21}\right),\nonumber
\\
\tilde{Y}^{23}&=&\frac{1}{2}\left(Y_1+Y_{2}+Y_{15}-Y_{14}+Y_{19}+Y_{21}-Y_{25}+Y_{26}\right),\nonumber
\\
\tilde{Y}^{24}&=&\frac{1}{2}\left(Y_{10}+Y_{9}-Y_{30}-Y_{31}+Y_6-Y_{7}-Y_{20}-Y_{24}\right),\nonumber
\\
\tilde{Y}^{25}&=&\frac{1}{2}\left(Y_2-Y_{1}-Y_{25}-Y_{26}-Y_{14}-Y_{15}+Y_{19}-Y_{21}\right),\nonumber
\\
\tilde{Y}^{34}&=&\frac{1}{2}\left(Y_8-Y_{4}-Y_{11}-Y_{28}+Y_{13}-Y_{32}+Y_{22}+Y_{23}\right),\nonumber
\\
\tilde{Y}^{35}&=&\frac{1}{2}\left(Y_{18}+Y_{17}-Y_{12}+Y_{27}-Y_{29}-Y_{16}-Y_{5}-Y_{3}\right),\nonumber
\\
\tilde{Y}^{45}&=&\frac{1}{2}\left(Y_8+Y_{4}-Y_{11}-Y_{28}-Y_{13}+Y_{32}-Y_{23}+Y_{22}\right).
\end{eqnarray}
This choice is of course not unique.
\\
\indent The scalar matrices for the moment maps are given by
\begin{eqnarray}
\mc{V}_{{\rm{a}}}^{ij,IJ}&=&-\frac{1}{6}\textrm{Tr}(L^{-1}J^{ij}LT^{IJ}),\nonumber
\\
\mc{V}_{{\rm{b}}}^{ij,IJ}&=&-\frac{1}{6}\textrm{Tr}(L^{-1}t^{ij}LT^{IJ}),\nonumber
\\
\mc{V}_{{\rm{a}}}^{ij,A}&=&\frac{1}{6}\textrm{Tr}(L^{-1}J^{ij}LY^{A}),\nonumber
\\
\mc{V}_{{\rm{b}}}^{ij,A}&=&\frac{1}{6}\textrm{Tr}(L^{-1}t^{ij}LY^{A})\label{V_map}
\end{eqnarray}
from which the T-tensor follows
\begin{eqnarray}
T^{IJ,KL}&=&g
\left(\mc{V}_{{\rm{a}}}^{ij,IJ}\mc{V}_{{\rm{b}}}^{ij,KL}+\mc{V}_{{\rm{b}}}^{ij,IJ}\mc{V}_{{\rm{a}}}^{ij,KL}\right)\nonumber
\\
T^{IJ,A}&=&g
\left(\mc{V}_{{\rm{a}}}^{ij,IJ}\mc{V}_{{\rm{b}}}^{ij,A}+\mc{V}_{{\rm{b}}}^{ij,IJ}\mc{V}_{{\rm{a}}}^{ij,A}\right)\label{T_ten}
\end{eqnarray}
Using these together with \eqref{A1}, \eqref{A2} and
\eqref{potential}, we can find the tensors $A_1^{IJ}$ and
$A^{IJ}_{2i}$ as well as the scalar potential.
\section{Non-semisimple gauging of $N=9$ gauged supergravity in three
dimensions}\label{N9_theory} We will consider $N=9$ gauged
supergravity in three dimensions. The corresponding scalar manifold
is given by the 16-dimensional $F_{4(-20)}/SO(9)$ coset space. Some
vacua of the compact and non-compact gaugings of this theory have
been studied in \cite{AP}. In this appendix, we will explore the
possibilities of non-semisimple gauge groups which are crucial for
embedding the theory in higher dimensions. Notice that the
construction of $E_6$ given in \cite{E6} is based on the $F_4$ group
given in \cite{F4}. We can simply remove the last 26 matrices $c_i$,
$i=53,\ldots, 78$ from $E_6$ to get the group $F_4$ generated by
$c_i$, $i=1,\ldots, 52$ as has been used in \cite{AP}. All 52
matrices are effectively $26\times 26$ matrices since all elements
in the last row and last column are zero.
\\
\indent The $SO(9)$ R-symmetry generators are $T^{IJ}$ in
\eqref{TIJ_gen} with $I,J=1,\ldots, 9$, and non-compact generators
are the first 16 generators of \eqref{non_compact_gen}, $Y^A$,
$A=1,\ldots, 16$. In the case of $F_{4(4)}/USp(6)\times SU(2)$ which
is a scalar manifold of $N=4$ theory studied in \cite{N4_gauging},
$SO(4)\ltimes \mathbf{T}^{6}$ can be gauged consistently with
supersymmetry by the embedding of $SO(4)\ltimes \mathbf{T}^{6}$ in
$SO(5,4)\subset F_{4(4)}$. In the present case, the embedding of
$SO(3)\ltimes \mathbf{T}^3$ in $USp(2,2)\subset USp(4,2)\times
SU(2)\subset F_{4(-20)}$ should be possible.
\\
\indent To identify generators of this group, we first consider the
$SO(4)\ltimes \mathbf{T}^6$ subgroup of the $SO(5)\ltimes
\mathbf{T}^{10}$ in section \ref{DW_solution}. Obviously, the
$SO(4)$ part is generated by $J^{ij}$, $i,j=1,\ldots, 4$. We then
consider $\tilde{Y}^{ij}$ with $i,j=1,\ldots, 4$. It can be verified
that by removing $Y_{17}$ to $Y_{32}$ form $\tilde{Y}^{ij}$, the
resulting generators, see appendix \ref{detail},
\begin{eqnarray}
\tilde{Y}^{12}&=&\frac{1}{2}\left(Y_3-Y_{12}+Y_5-Y_{16}\right),\nonumber
\\
\tilde{Y}^{13}&=&\frac{1}{2}\left(Y_2+Y_{14}-Y_1+Y_{15}\right),\nonumber
\\
\tilde{Y}^{14}&=&\frac{1}{2}\left(Y_{10}-Y_{7}-Y_{6}-Y_{30}-Y_9\right),\nonumber
\\
\tilde{Y}^{23}&=&\frac{1}{2}\left(Y_1+Y_{2}+Y_{15}-Y_{14}\right),\nonumber
\\
\tilde{Y}^{24}&=&\frac{1}{2}\left(Y_{10}+Y_{9}+Y_6-Y_{7}\right),\nonumber
\\
\tilde{Y}^{34}&=&\frac{1}{2}\left(Y_8-Y_{4}-Y_{11}+Y_{13}\right)
\end{eqnarray}
still transform in the adjoint representation of $SO(4)$. It turns
out that when combined into $t^{ij}$, the resulting generators do
not commute. Therefore, it is not possible to find $SO(4)\ltimes
\mathbf{T}^6$ subgroup of $F_{4(-20)}$. On the other hand, we can
form two $SU(2)_{\pm}$ subgroups from these generators by
introducing the self-dual and anti-self-dual $SO(4)$ generators
\begin{eqnarray}
J_+^1&=&J^{12}+J^{34},\qquad J^2_+=J^{13}-J^{24},\qquad
J^3_+=J^{14}+J^{23},\nonumber \\
t_+^1&=&t^{12}+t^{34},\qquad t^2_+=t^{13}-t^{24},\qquad
t^3_+=t^{14}+t^{23}
\end{eqnarray}
and
\begin{eqnarray}
J_-^1&=&J^{12}-J^{34},\qquad J^2_-=J^{13}+J^{24},\qquad
J^3_-=J^{14}-J^{23},\nonumber \\
t_-^1&=&t^{12}-t^{34},\qquad t^2_-=t^{13}+t^{24},\qquad
t^3_-=t^{14}-t^{23}\, .
\end{eqnarray}
It can be readily verified that each set of generators forms
$SO(3)\ltimes \mathbf{T}^3\sim SU(2)\ltimes \mathbf{T}^3$ algebra
but generators $t^a_\pm$ from the two sets do not commute with
eachMo other. Although this subgroup can be embedded in
$F_{4(-20)}$, it is not admissible namely it cannot be gauged in a
way that is consistent with supersymmetry. Embedding in higher
dimensions aside, it seems to be difficult (if possible) to find
non-semisimple gaugings of the $N=9$ theory.

\end{document}